\documentclass[prb,twocolumn,showpacs]{revtex4}

\usepackage{graphicx}%
\usepackage{dcolumn}
\usepackage{amsmath}

\makeatletter
\def\btt#1{\texttt{\@backslashchar#1}}%
\DeclareRobustCommand\bblash{\btt{\@backslashchar}}%
\makeatother

\begin{document}

\preprint{}

\title[Charge ordering in $\theta$-RbZn]{
Charge ordering in $\theta$-(BEDT-TTF)$_2$RbZn(SCN)$_4$: \\
Cooperative effects of electron correlations and lattice distortions
}

\author{Satoshi Miyashita$^{1,2}$}
 \email{satoshi@ims.ac.jp}
\author{Kenji Yonemitsu$^{2,3}$}%
\affiliation{%
$^1$Institute for Materials Research, Tohoku University, 
Sendai 980-8577, Japan \\
$^2$Institute for Molecular Science, 
Okazaki 444-8585, Japan \\
$^3$Department of Functional Molecular Science, Graduate University 
for Advanced Studies, Okazaki 444-8585, Japan 
}%


\date{\today}

\begin{abstract}
Combined effects of electron correlations and lattice distortions 
are investigated on the charge ordering 
in $\theta$-(BEDT-TTF)$_2$RbZn(SCN)$_4$ theoretically 
in a two-dimensional 3/4-filled extended Hubbard model with 
electron-lattice couplings. 
It is known that this material undergoes a phase transition from 
a high-symmetry metallic state to a low-symmetry insulating state with 
a horizontal-stripe charge order (CO) by lowering temperature. 
By means of the exact-diagonalization method, 
we show that electron-phonon interactions are crucial to stabilize 
the horizontal-stripe CO and to realize the low-symmetry crystal structure.
\end{abstract}


\maketitle


\section{Introduction}
Two-dimensional (2D) strongly correlated electron systems
have attracted much attention in condensed matter physics. 
For bis(ethylenedithio)-tetrathiafulvalene (BEDT-TTF) salts, 
which are 2D strongly correlated organic conductors 
with various molecular arrangements,\cite{Ishiguro,Williams} 
many theoretical scientists use the extended Hubbard model 
in order to understand the role played by long-range Coulomb interactions. 
It should be noted that not only electron-electron interactions 
but also electron-phonon couplings can be significant 
because the crystal structure is often altered at a phase transition. 
These materials exhibit many interesting phenomena such as 
metal-insulator (M-I) transitions with charge ordering, 
the realization of a spin liquid state,\cite{Shimizu} 
superconductivity,\cite{Kanoda} and so on. 
In particular, 
transitions to states with a charge order (CO) have been investigated 
theoretically\cite{Kino,Seo1,Mazumdar, Clay,SeoReview} 
and experimentally.\cite{HMori1,HMori2} 

\begin{figure}[thb]
\includegraphics[width=80mm]{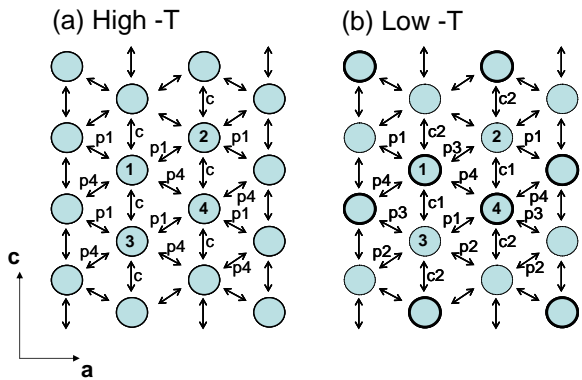}
\caption{(Color online) 
Anisotropic triangular lattice for $\theta$-RbZn salt: 
(a) high-symmetry structure at high temperature, and 
(b) low-symmetry structure at low temperature 
with horizontal-stripe CO along $t_{p4}$ bonds. 
The thin and thick circles represent the hole-poor and the hole-rich sites, 
respectively.
}
\label{modelF}
\end{figure}

$\theta$-(BEDT-TTF)$_2$RbZn(SCN)$_4$ 
(called $\theta$-RbZn for simplicity hereafter) 
is a representative compound which undergoes a CO transition accompanied with 
a structural deformation by lowering temperature.\cite{HMori1,HMori2} 
The ground state of the $\theta$-RbZn salt is an insulator with the 
horizontal-stripe CO along $t_{p4}$ bonds (HCO-$t_{p4}$) 
shown in Fig. \ref{modelF}(b), where sites $1$ and $4$ are hole-rich, 
as confirmed by the X-ray structural analysis,\cite{Watanabe} 
$^{13}$C-NMR measurements\cite{Miyagawa,Chiba} and 
polarized Raman and IR spectroscopy.\cite{HTajima,Yamamoto} 
The importance of long-range electron-electron interactions is well recognized 
and the mechanism for stabilizing the HCO-$t_{p4}$ has been argued mainly 
on the basis of the low-symmetry structure. 
Because the lattice distortions are coupled with the electron system, 
giving rise to the first-order transition sensitive to the crystal structure, 
electron-phonon interactions are also important.

Quite recently, Iwai $et$ $al$. have observed photoinduced melting of CO in 
the $\theta$-RbZn salt and 
in $\alpha$-(BEDT-TTF)$_2$I$_3$ (called $\alpha$-I$_3$ 
for simplicity hereafter) by femtosecond reflection spectroscopy.\cite{Iwai} 
The $\theta$-RbZn and $\alpha$-I$_3$ salts show large and small 
molecular rearrangements, respectively, at the M-I transition. 
Their photoinduced dynamics are qualitatively different: 
the $\theta$-RbZn salt shows local melting of CO and ultrafast recovery of CO 
irrespective of temperature and excitation intensity, 
while the $\alpha$-I$_3$ salt shows critical slowing down. 
Thus, it is important to show how electron-phonon interactions are significant 
in the $\theta$-RbZn salt. 

This paper is organized as follows. After a brief explanation of the
model in the next section,  we present the exact-diagonalization results
in Sec.\ref{numerical} for the hole-hole correlation functions, 
the hole densities, and the modulations of transfer integrals, 
from which we propose that 
the low-symmetry structure of the $\theta$-RbZn salt at low temperature 
is reproduced 
by introducing electron-lattice couplings in the model based on 
the high-symmetry structure at high temperature. 
Then, in Sec.\ref{discussion}, 
we discuss the numerical results on the basis of a perturbation theory 
from the strong-coupling limit. 
A brief summary is given in Sec.\ref{summary}. 
%
\section{Model}
%
%
We start with the following extended Hubbard model at 3/4-filling 
with electron-phonon couplings of transfer-modulation type, 
\begin{eqnarray}
 {\cal H} &=& \sum_{\langle i,j \rangle}\sum_{\mu=c,p1,p4} 
              \left[
               t_{i,j} \pm \alpha_\mu u_\mu
              \right] 
              c^\dag_{i,\sigma} c_{j,\sigma}
           +  U \sum_{i} n_{i,\uparrow}n_{i,\downarrow}
\nonumber \\
   &&      +  V_c \sum_{\langle i,j \rangle_c} n_i n_j
           +  V_p \sum_{\langle i,j \rangle_p} n_i n_j
           +  \sum_{\mu=c,p1,p4} \frac{K_\mu}{2} u_\mu^2 \;, 
\label{model}
\end{eqnarray}
where $c^\dag_{i,\sigma}$ creates an electron with spin $\sigma$ at site $i$, 
$n_{i,\sigma}=c^\dag_{i,\sigma} c_{i,\sigma}$, and 
$n_{i}=\sum_{\sigma}n_{i,\sigma}$. 
$t_{i,j}=t_c^{\rm HT}$, $t_{p1}^{\rm HT}$ or $t_{p4}^{\rm HT}$ 
is the transfer integral for the bond between 
the $i$-th site and its nearest-neighbor $j$-th site 
along the $c$, $p1$ or $p4$ bond. 
$\langle i,j \rangle_c$ and $\langle i,j \rangle_p$ denote 
the nearest-neighbor pairs $i$ and $j$ 
along the $c$ bond and the $p$ bond, respectively. 
$U$ represents the on-site Coulomb interaction, and 
$V_{c(p)}$ is the intersite Coulomb interaction between the $i$-th site and 
the $j$-th site on the $c$($p$)-bond. 
$u_i$ is the $i$-th molecular translation or rotation 
explained later from the equilibrium position in the high temperature phase. 
$\alpha_\mu$ and $K_\mu$ are the electron-phonon coupling strength and 
the elastic coefficient, respectively. 
For simplicity, we perform variable transformations as 
\begin{eqnarray}
       \alpha_\mu u_{\mu} = y_{\mu} \;, \ \ 
       \frac{\alpha_\mu^2}{K_{\mu}} = s_{\mu} \;. 
\label{transformation}
\end{eqnarray}
%
\section{Numerical results}
\label{numerical}
We show energy values in the unit of electron volt (eV) 
in the following. 
For electron-phonon couplings, 
we consider three kinds of molecular displacements: 
translations in the $c$-direction ($u_c$), 
those in the $a$-direction ($u_{p1}$), 
and rotations ($u_{p4}$) as discussed in terms of elevation angles 
by Watanabe $et$ $al$.\cite{Watanabe} 
We suppose that translations in the $c$($a$)-direction 
contribute to modulations of the transfer integrals 
on the $c1$-($p1$-) and $c2$-($p3$-)bonds. 
We also assume that molecular rotations produce differences between the 
transfer integral on the $p2$-bond and that on the $p4$-bond. 
Then, the modulated transfer integral on each bond reads 
\begin{eqnarray}
\begin{array}{ll}
t_{c1} = t_{c}^{\rm HT} - \alpha_{c} u_{c} \;, &
t_{c2} = t_{c}^{\rm HT} + \alpha_{c} u_{c} \;, \\
t_{p1} = t_{p1}^{\rm HT} + \alpha_{p1} u_{p1} \;, &
t_{p3} = t_{p1}^{\rm HT} - \alpha_{p1} u_{p1} \;, \\
t_{p2} = t_{p4}^{\rm HT} + \alpha_{p4} u_{p4} \;, &
t_{p4} = t_{p4}^{\rm HT} - \alpha_{p4} u_{p4} \;. 
\end{array}
\label{formation}
\end{eqnarray}
The signs here are so determined that $y_\mu>0$ corresponds to a deviation 
from the high-temperature crystal structure toward 
the low-temperature one of the $\theta$-RbZn salt. 

In this section, 
we adopt $U=0.7$ and $V_c$, $V_p \leq U/2$.\cite{Ducasse,Imamura,TMori1} 
Typical values for transfer integrals $t_\mu$ in BEDT-TTF salts 
are estimated from the extended H\"uckel calculation.\cite{TMori2}
We suppose here from the experimental data\cite{Watanabe} that 
$t_{c}^{\rm HT}=0.035$, 
$t_{p1}^{\rm HT}=0.095$, and $t_{p4}^{\rm HT}=-0.095$ 
correspond to the high-symmetry structure at high temperature, 
where the $p1$ and $p4$ bonds are equivalent. 

We use the exact-diagonalization method for electrons, 
regard phonons as classical variables, 
and apply the Hellmann-Feynman theorem 
to impose the self-consistency condition on the phonons 
($y_c$, $y_{p1}$, $y_{p4}$) determined by 
\begin{eqnarray}
      \langle
       \frac{\partial {\cal H}}{\partial y_{\mu}}
      \rangle
       = 0 \;. 
\label{HF-theorem}
\end{eqnarray}
We vary mainly coupling strengths, $s_c$, $s_{p1}$, and $s_{p4}$. 
%
\subsection{Case without electron-phonon couplings}
%
\begin{figure*}[thb]
\includegraphics[width=150mm]{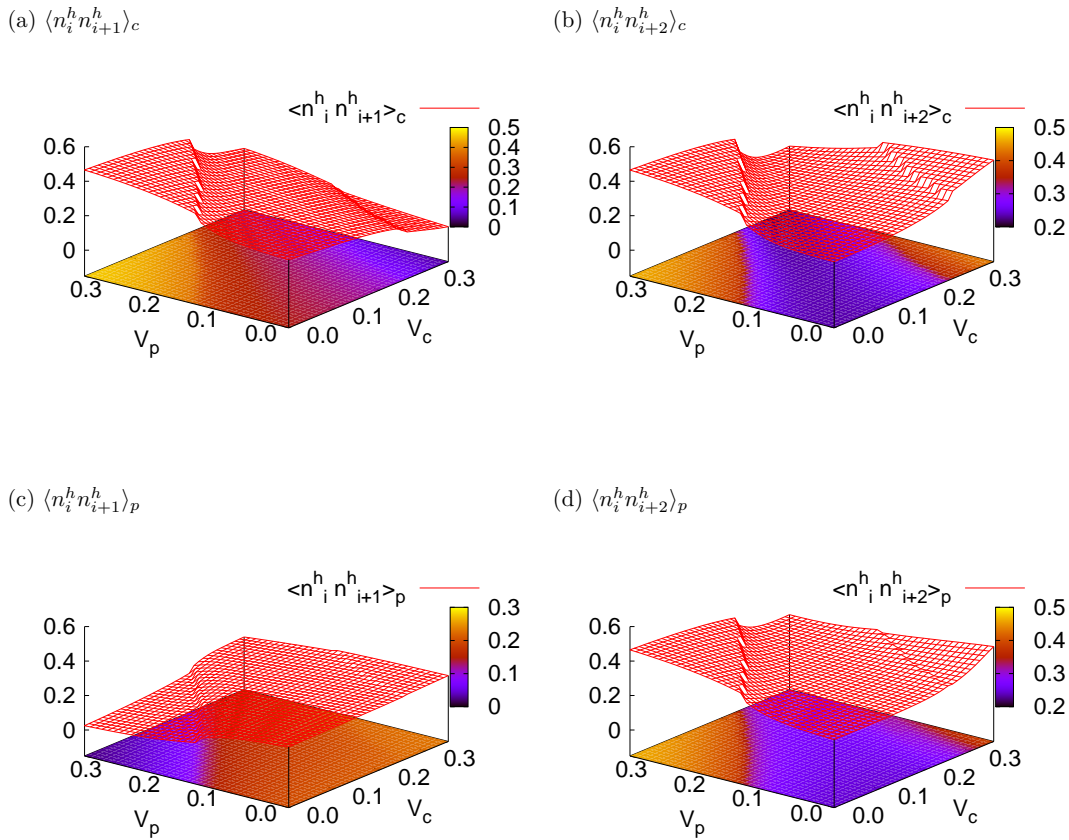}
\caption{
Hole-hole correlation functions as a function of $V_c$ and $V_p$ 
for $U=0.7$ without electron-phonon couplings: 
(a) $\langle n^h_i n^h_{i+1} \rangle_c$ and 
(b) $\langle n^h_i n^h_{i+2} \rangle_c$ 
are correlation functions along vertical bonds. 
(c) $\langle n^h_i n^h_{i+1} \rangle_p$ and 
(d) $\langle n^h_i n^h_{i+2} \rangle_p$ 
are correlation functions along diagonal bonds. 
}
\label{corre}
\end{figure*}
At the beginning, we clarify the ground-state properties 
of the model (\ref{model}) 
in the absence of lattice distortions. 
In this case, the exact-diagonalization studies 
of the 4$\times$4-site cluster show that the hole densities are uniform 
($\langle n^h_i \rangle=1-\langle n_i \rangle=0.5$) 
in any combination of $V_c$ and $V_p$ (not shown). 
Degenerate CO states are mixed owing to the finite-size effect. 
Then, we have calculated correlation functions in this cluster.

Figure \ref{corre} shows the hole-hole correlation functions 
without electron-phonon couplings on the $V_c-V_p$ plane. 
$\langle n^h_i n^h_{i+1(i+2)} \rangle_\mu$ denotes 
the hole-hole correlation function along the $\mu$-bond 
between the $i$-th and its neighbor (its second-neighbor) site. 
$\langle n^h_i n^h_{j} \rangle_\mu$ being nearly equal to $0.5$ 
means that 
$(\langle n^h_i \rangle, \langle n^h_{j} \rangle)=(1-\delta, 1-\delta)$ 
is equally mixed with $(\delta, \delta)$ with small $\delta$. 
On the other hand, 
very small $\langle n^h_i n^h_j \rangle_\mu$ represents that 
$(\langle n^h_i \rangle, \langle n^h_j \rangle)=(1-\delta, \delta)$ 
is mixed with $(\delta, 1-\delta)$. 
Namely, 
$\langle n^h_i n^h_j \rangle$ is close to 0 (0.5) when 
the charge disproportionation between the $i$-th and the $j$-th sites 
is large (small). 

For large $V_c$, e.g., for ($V_c$, $V_p$)=(0.35, 0.01), 
$\langle n^h_i n^h_{i+1} \rangle_c \sim 0.05$,
$\langle n^h_i n^h_{i+2} \rangle_c \sim 0.44$,
$\langle n^h_i n^h_{i+1} \rangle_p \sim 0.23$, and 
$\langle n^h_i n^h_{i+2} \rangle_p \sim 0.39$. 
This result means that the diagonal-stripe CO is the largest at this point 
among all hole-hole correlations. 
On the other hand, 
for small $V_c$, e.g., for ($V_c$, $V_p$)=(0.01, 0.35), 
$\langle n^h_i n^h_{i+1} \rangle_c \sim 0.46$,
$\langle n^h_i n^h_{i+2} \rangle_c \sim 0.46$,
$\langle n^h_i n^h_{i+1} \rangle_p \sim 0.03$, and 
$\langle n^h_i n^h_{i+2} \rangle_p \sim 0.46$. 
Here, the vertical-stripe CO is the largest correlation. 
In the ground state with comparable $V_c$ and $V_p$, 
the diagonal- and the vertical-stripe COs coexist. 
%
\begin{figure}[thb]
\includegraphics[width=70mm]{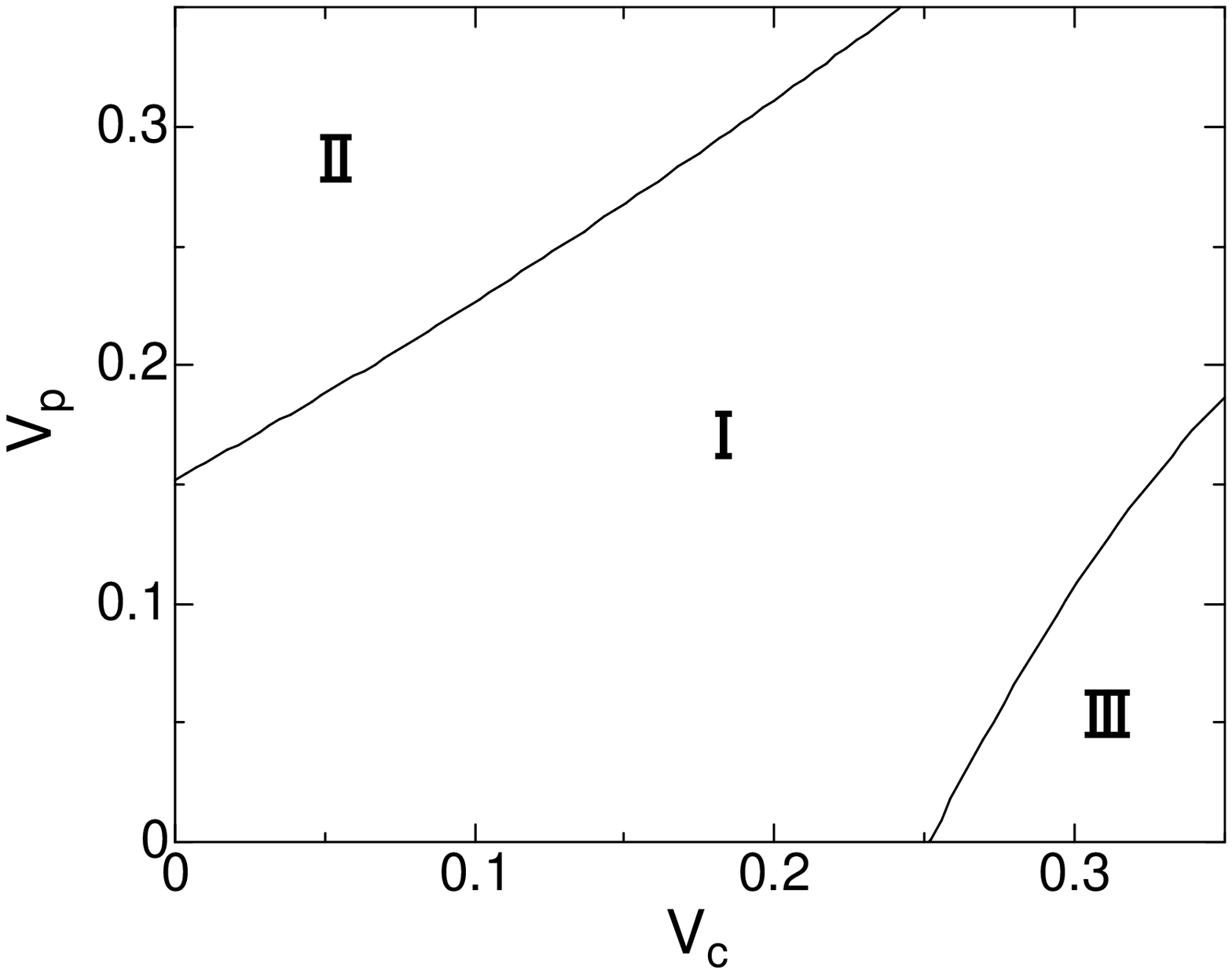}
\caption{
Phase diagram for $U=0.7$ without electron-phonon couplings. 
All phases are uniform, but the dominant hole-hole correlations are different. 
The vertical-stripe CO and the diagonal-stripe CO coexist in phase I. 
The vertical-stripe CO is the largest correlation in phase I$\!$I, 
while the diagonal-stripe CO is the largest correlation in phase I$\!$I$\!$I. 
}
\label{phase-diagram}
\end{figure}
%
Regarding the steep variation in the contour plots of the 
correlation functions in Fig. \ref{corre} as a phase boundary, 
we obtain the ground-state phase diagram on the $V_c$-$V_p$ plane 
in Fig. \ref{phase-diagram}. 
The phase diagram consists of three uniform phases 
with different hole-hole correlations; 
the coexistent phase (I), 
the phase with dominant vertical-stripe CO correlation (I$\!$I), and 
the phase with dominant diagonal-stripe CO correlation (I$\!$I$\!$I). 
This phase diagram is consistent with the previous work.\cite{Merino} 
The most important thing here is that 
we cannot find the horizontal-stripe CO phase 
in any combination of $V_c$ and $V_p$, which is observed experimentally 
in the $\theta$-RbZn salt at low temperature. 

\begin{figure}[thb]
\includegraphics[width=70mm]{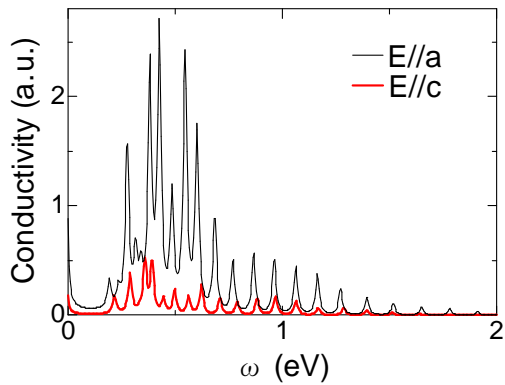}
\caption{(Color online) 
Optical conductivity spectra of the model (\ref{model}) 
on the 4$\times$4-site cluster for electric fields, 
$E{\parallel}a$ (thin line) and $E{\parallel}c$ (thick line). 
We use $U=0.7$, $V_c=0.31$ and $V_p=0.27$. 
The $\delta$-functions appearing in the continued fraction expansion 
are broadened with width $\eta$=0.01. 
}
\label{opt}
\end{figure}

Below we mainly use $V_c=0.31$, $V_p=0.27$ 
in phase I of Fig. \ref{phase-diagram}, where this ratio of 
$V_c/V_p$ is regarded as appropriate for the $\theta$-RbZn salt.\cite{TMori} 
To check whether these values themselves are appropriate, 
we have calculated the optical conductivity spectra 
by means of the continued fraction expansion to  show them in Fig. \ref{opt}.
\cite{EDagotto} 
It is found that for both polarizations $E$ 
they have a broad peak around $\omega{\sim}0.4$ eV and 
the conductivity for $E{\parallel}a$ is larger than that for $E{\parallel}c$, 
which are consistent with the experimental findings.\cite{HTajima2}
Therefore, these interaction strengths are quite reasonable. 
%
%
\subsection{Case with electron-phonon couplings}
%
\begin{figure*}[thb]
\includegraphics[width=120mm]{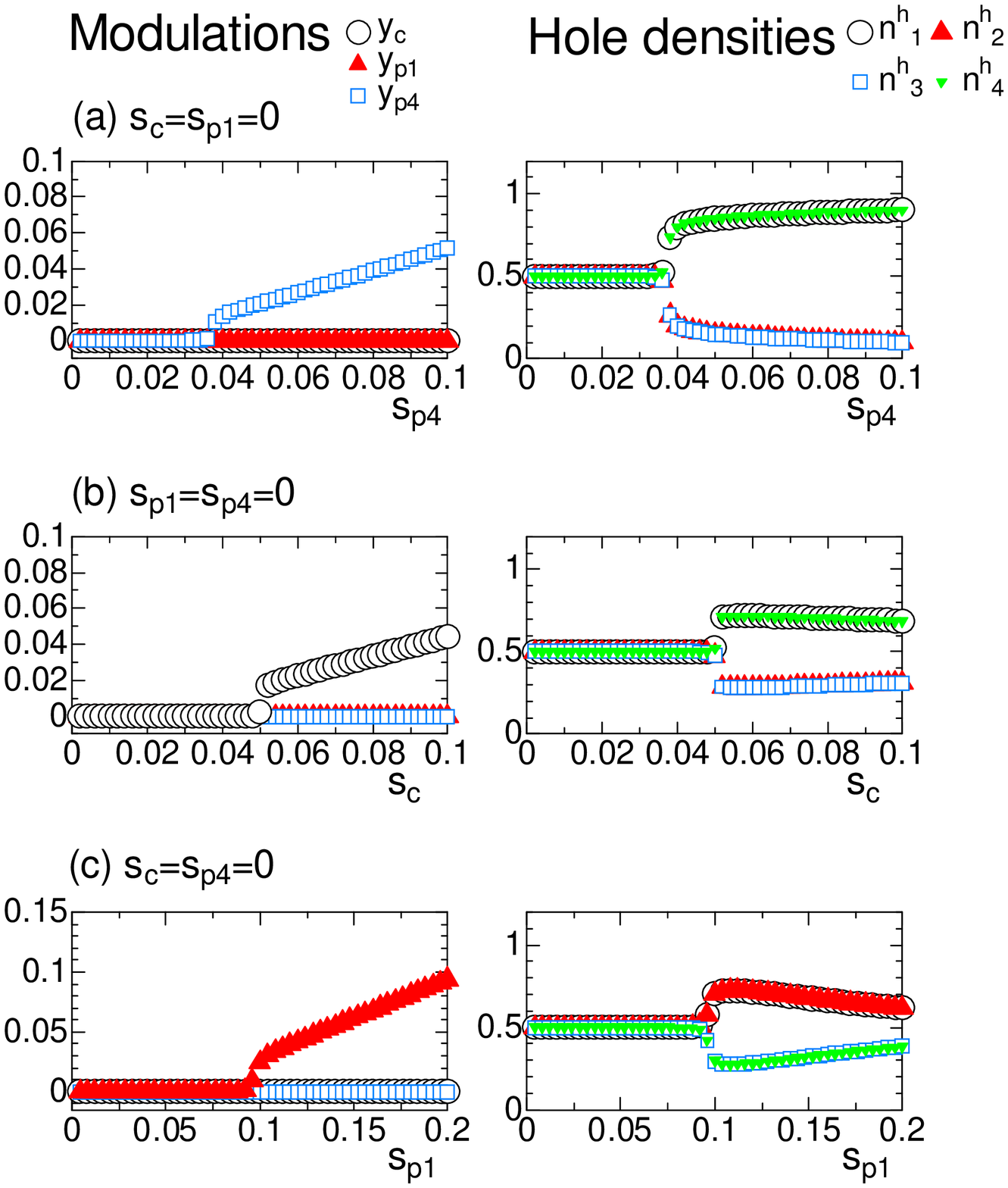}
\caption{(Color online) 
(a) $s_{p4}$-dependence of modulations of transfer integrals 
(left) and hole densities (right) for $s_{c}$=$s_{p1}$=0, 
(b) their $s_{c}$-dependence for $s_{p1}$=$s_{p4}$=0, and 
(c) their $s_{p1}$-dependence for $s_{c}$=$s_{p4}$=0, 
with $U=0.7$, $V_c=0.31$ and $V_p=0.27$. 
}
\label{all}
\end{figure*}
%
Each of the three kinds of electron-phonon couplings is studied 
at the particular point of $V_c=0.31$ and $V_p=0.27$ 
by the exact diagonalization. 
Because we need to obtain the displacements consistently with bond densities, 
we used the smaller system of the 12 sites hereafter 
and compared some results with those of the 16-site cluster. 
The conclusion is found to be unchanged and, in fact, 
understood from the perturbation theory from the strong-coupling limit, 
as discussed later. 

First, we see the effect of molecular rotations ($s_{p4}$). 
The $s_{p4}$-dependence of the modulations of transfer integrals 
and the hole densities is shown in Fig. \ref{all}(a) 
with $s_c=s_{p1}=0$ fixed. 
For small $s_{p4}$, 
as all the phonons are undistorted and the hole densities are uniform 
($\langle n^h_i \rangle=0.5$), 
the ground state remains the coexistent state of 
the vertical-stripe CO and the diagonal-stripe CO. 
With increasing $s_{p4}$, the system discontinuously changes 
at a critical point $s_{p4}^{\rm cr}\sim0.04$, 
to the CO state of broken symmetry. 
This CO pattern is the HCO-$t_{p4}$ 
($\langle n^h_1 \rangle$, $\langle n^h_4 \rangle$ $\gg$ 
 $\langle n^h_2 \rangle$, $\langle n^h_3 \rangle$), 
which agrees with the experimental findings.\cite{Watanabe} 
Therefore, the coupling with molecular rotations ($s_{p4}$) 
plays an important role to realize this HCO state. 
It is noted that $y_{p4}$ increases almost linearly 
after the critical point where the HCO-$t_{p4}$ state is stable. 
This discontinuous change at the critical point would be caused by 
the finite-size effect: 
the undistorted state is stabilized by quantum tunneling between 
different CO states, but its energy gain would vanish 
in the thermodynamic limit. 
In fact, $y_{p4}$ rises up linearly by infinitesimal $s_{p4}$ 
in mean-field calculations,\cite{Tanaka} 
so we guess that the critical point approaches zero 
in the thermodynamic limit. 

Next, we consider the effect of translations in the $c$-direction ($s_c$). 
For large $s_c$, 
they also stabilize the HCO-$t_{p4}$ state as shown in Fig. \ref{all}(b). 
In the case of $s_c$, the charge disproportionation between 
the hole-rich sites ($\langle n^h_1 \rangle$, $\langle n^h_4 \rangle$) and 
the hole-poor sites ($\langle n^h_2 \rangle$, $\langle n^h_3 \rangle$) 
is smaller than the case of $s_{p4}$, but 
the translations in the $c$-direction further stabilize the HCO-$t_{p4}$. 

The situation regarding $s_{p1}$ is quite different from 
the cases of $s_{p4}$ and $s_{c}$. 
The $s_{p1}$-dependence of the transfer modulations and 
the hole densities at four sites is shown in Fig. \ref{all}(c). 
In the presence of only $y_{p1}$, which represents 
translations in the $a$-direction, 
the CO pattern is still a horizontal-type but the holes are localized 
on the $t_{p1}$ and $t_{p3}$ bonds. 
Thus, the experimental findings are not reproduced 
if only $s_{p1}$ exists. 

From Figs. \ref{all}(a)-\ref{all}(c), 
we conclude that the effects of molecular translations 
in the $c$-direction ($s_c$) and molecular rotations ($s_{p4}$) are stronger 
than that of translations in the $a$-direction ($s_{p1}$) in this salt. 
\begin{figure*}[thb]
\includegraphics[width=120mm]{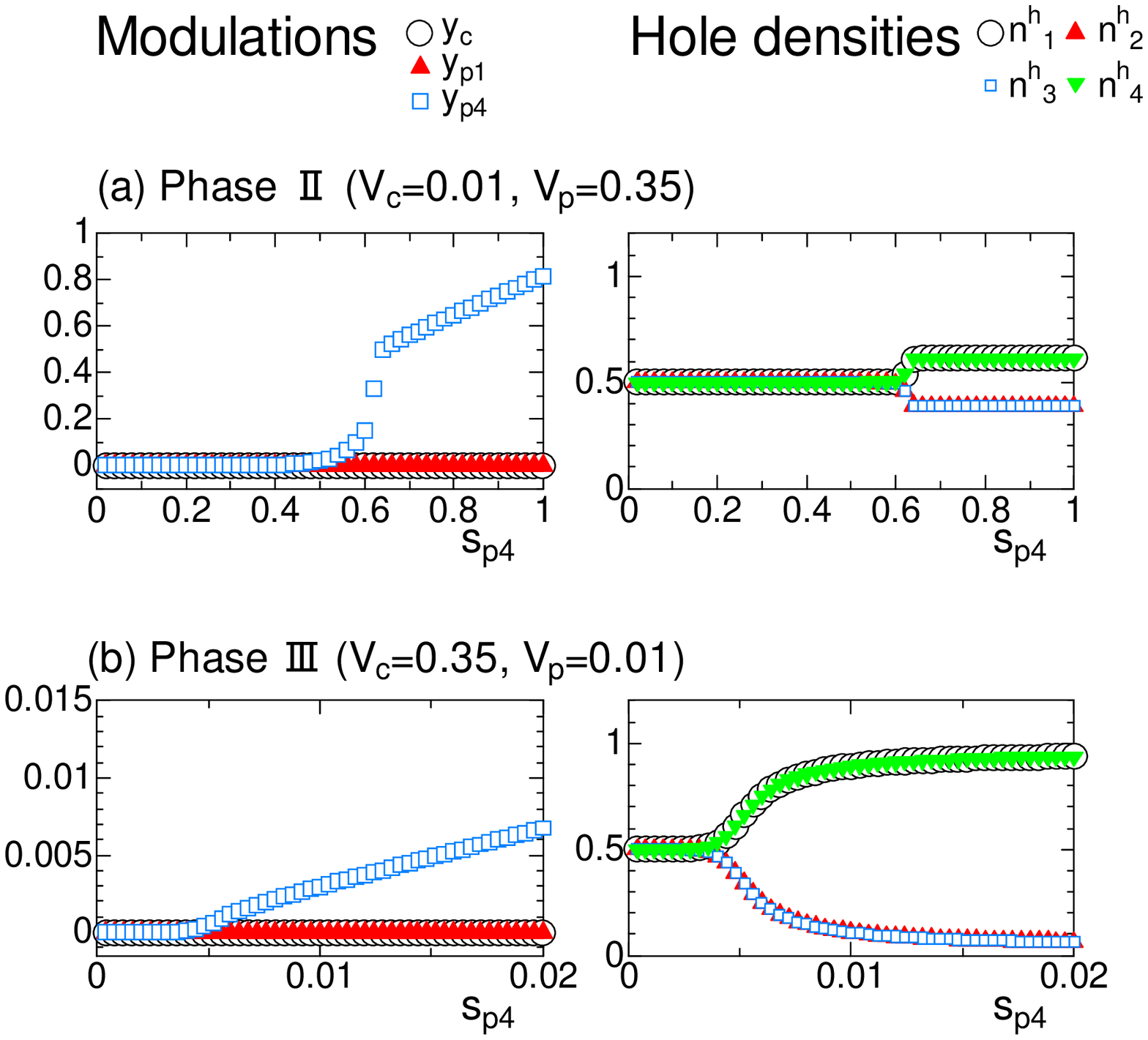}
\caption{(Color online) 
$s_{p4}$-dependence of modulations of transfer integrals 
(left) and hole densities (right), 
(a) in phase I$\!$I ($V_c=0.01$, $V_p=0.35$), and 
(b) in phase I$\!$I$\!$I ($V_c=0.35$, $V_p=0.01$), 
for U=0.7 with $s_{c}$=$s_{p1}$=0 fixed. 
}
\label{variousV}
\end{figure*}
%
We also come to the same conclusion in the other phases 
of Fig. \ref{phase-diagram}. 
In Fig. \ref{variousV}, 
we perform the same calculations (except for the parameters shown) 
as in Fig. \ref{all}(a) for phases I$\!$I and I$\!$I$\!$I. 
The $s_{p4}$-induced HCO-$t_{p4}$ state is stable 
in both of phases I$\!$I and I$\!$I$\!$I when $s_{p4}$ is large enough. 
Therefore, this characteristic phonon-induced feature is widely obtained 
on the $V_c$-$V_p$ plane, including $V_c=V_p=0$ 
(irrespective of hole-hole correlations in the uniform ground state 
 without electron-phonon couplings). 

\begin{figure}[thb]
\includegraphics[width=80mm]{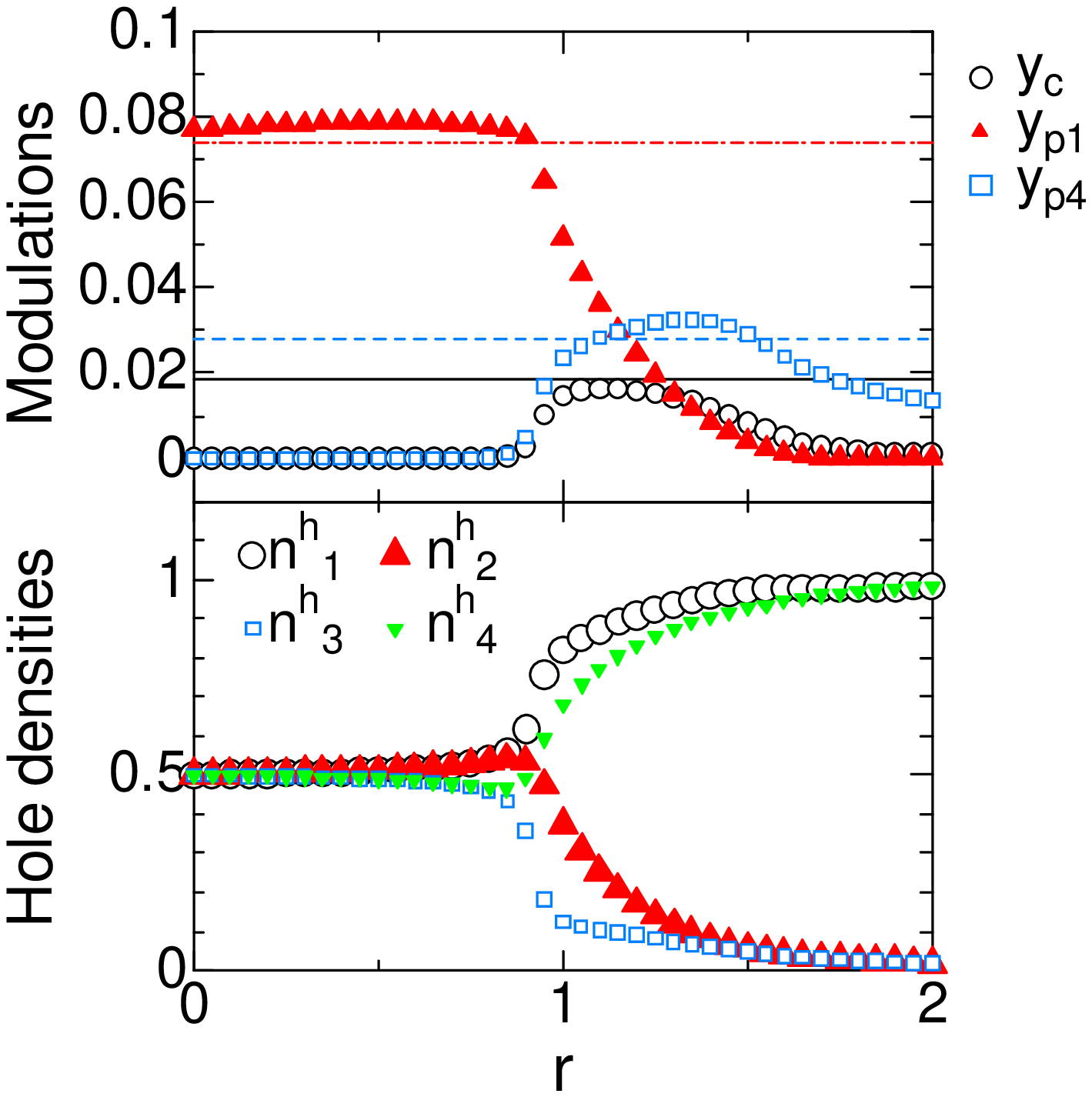}
\caption{(Color online) 
Three kinds of transfer modulations and hole densities at four sites 
for $U=0.7$, $V_c=0.31r$ and $V_p=0.27r$. 
Coupling strengths are fixed at $s_{c}=0.05$, $s_{p1}=0.17$ and $s_{p4}=0.06$. The solid (dashed-dotted) line in the upper panel represents 
the magnitude of the molecular translations in the $c$($a$)-direction 
in the low-symmetry structure of $\theta$-RbZn 
at low temperature.\cite{Watanabe}
The dashed line represents the magnitude of the molecular rotations 
in this structure. 
}
\label{strng-cplng}
\end{figure}

Finally, 
we simultaneously take the three kinds of displacements into account. 
Considering a situation with thermal- or pressure-induced expansion or 
compression, we vary intersite Coulomb interactions. 
In Fig. \ref{strng-cplng}, we show the modulations of transfer integrals 
and the hole densities as a function of nearest-neighbor Coulomb interactions. 
For small $r$, only $y_{p1}$ is finite with $y_c$ and $y_{p4}$ being zero. 
This situation is quite similar to Fig. \ref{all}(c), and hence 
the most stable state is the HCO-$t_{p1}$\&$t_{p3}$. 
As $r$ approaches unity, which is a realistic value for $\theta$-RbZn, 
the experimentally observed HCO-$t_{p4}$ state 
becomes more stable than the HCO-$t_{p1}$\&$t_{p3}$ 
by gradually increasing $y_c$ and $y_{p4}$ and rapidly decreasing $y_{p1}$. 
In the vicinity of $r=1$ for $s_c=0.05$, $s_{p1}=0.17$, and $s_{p4}=0.06$, 
$y_c$ and $y_{p4}$ approximately correspond to 
the low-symmetry structure of $\theta$-RbZn at low temperature. 

From the X-ray structure analysis,\cite{Watanabe} the molecular translations 
in the $c$- and $a$-directions are found to be 
$u_c \sim 0.13 {\rm \AA}$ and $u_{p1} \sim 0.15 {\rm \AA}$, 
and the corresponding distortions are 
$y_c = 0.0185 {\rm eV}$  and $y_{p1} = 0.0735 {\rm eV}$. 
They give 
$\alpha_c \sim 0.14 {\rm eV}/{\rm \AA}$ and 
$\alpha_{p1} \sim 0.47 {\rm eV}/{\rm \AA}$. 
The present coupling strengths 
$s_c = 0.05 {\rm eV}$ and $s_{p1} = 0.17 {\rm eV}$
lead to 
$K_c \sim 0.38 {\rm eV}/{\rm \AA^2}$ and 
$K_{p1} \sim 1.32 {\rm eV}/{\rm \AA^2}$. 
A rough estimation of phonon frequencies by 
$\omega_\mu=\sqrt{K_\mu/m}$ with $m$ being the reduced mass of 
two BEDT-TTF molecules shows 
$\omega_c \sim 23 {\rm cm^{-1}}$ and $\omega_{p1} \sim 43 {\rm cm^{-1}}$, 
which are quite reasonable values.\cite{Iwai}

As $r$ further increases beyond two, 
all modulations approach zero. 
In particular, $y_{p1}$ decreases monotonically. 
We explain these behaviors with a perturbation theory 
from the strong-coupling limit in the following section. 
Thus, electron-phonon interactions are crucial to stabilize 
the horizontal-stripe CO and to realize the low-symmetry structure 
of $\theta$-RbZn at low temperature. 
All these results are consistent with mean-field studies 
of the same model.\cite{Tanaka}
%
\section{Discussion}
\label{discussion}
In this section, we consider why the HCO-$t_{p4}$ is the most stabilized 
by the electron-phonon couplings. 
It is easily understood in the strong-coupling limit. 
We assume that, in this limit, the holes are perfectly localized 
on the $t_{p4}$ bonds, namely 
$\langle n^h_1 \rangle = \langle n^h_4 \rangle =1$ 
as shown in Fig. \ref{modelF}(b). 
In the ordinary perturbation theory, 
the second- and third-order contributions to the energy per site read 
\begin{eqnarray}
E_2 &=& - \frac{ {t_{c1}}^2+{t_{c2}}^2 }{2 V_c }
        - \frac{ {t_{p1}}^2+{t_{p3}}^2 }{2 (2V_c-V_p) }
\nonumber \\
    &&  
        + \frac{ 2{t_{p4}}^2 }{ U-V_p }
          \left\{
           \langle {\bf S}_i \cdot {\bf S}_j \rangle - \frac{1}{4}
          \right\} \;, 
\label{2nd-perturb}
\\
E_3 &=& - \frac{ (t_{p1}+t_{p3}) t_{c2} t_{p2}}{ V_c(2V_c-V_p) }
         + \frac{ 2(t_{p1}+t_{p3}) t_{c1} t_{p4}}{ V_c(2V_c-V_p) }
          \left\{
           \langle {\bf S}_i \cdot {\bf S}_j \rangle + \frac{1}{4}
          \right\}
\nonumber \\
    &&  + \frac{ 2(t_{p1}+t_{p3}) t_{c1} t_{p4}}{ V_c(U-V_p) }
          \left\{
           \langle {\bf S}_i \cdot {\bf S}_j \rangle - \frac{1}{4}
          \right\}
\nonumber \\
    &&  
         + \frac{ 2(t_{p1}+t_{p3}) t_{c1} t_{p4}}{ (2V_c-V_p)(U-V_p) }
          \left\{
           \langle {\bf S}_i \cdot {\bf S}_j \rangle - \frac{1}{4}
          \right\} \;, 
\label{3rd-perturb}
\end{eqnarray}
where ${\bf S}_i$ is the $S=1/2$ spin operator at the $i$-th site 
on the $t_{p4}$ bond. 
In the strong-coupling limit, the one-dimensional half-filled chain is 
formed along the $t_{p4}$ bonds and 
$\langle {\bf S}_i \cdot {\bf S}_{i+1} \rangle 
= - \ln 2 + 1/4 \simeq -0.443$ is the exact ground-state energy 
of the isotropic $S=1/2$ Heisenberg chain. 
According to the third term in Eq. (\ref{2nd-perturb}), 
there is energy gain from spin fluctuations, 
so that the ground-state energy of the HCO-$t_{p4}$ state 
is further lowered by increasing $|t_{p4}|$. 
On the other hand, because all terms in Eq.~(\ref{3rd-perturb}) 
are positive, $E_3$ represents energy loss. 
When $|t_{p4}|$ becomes large, $|t_{c1}|$ becomes small to reduce 
the energy loss. 
From the relations (\ref{formation}), $|t_{c2}|$ becomes large and 
$|t_{p2}|$ becomes small. 

Equations (\ref{2nd-perturb}) and (\ref{3rd-perturb}) can be rewritten as 
\begin{eqnarray}
 E_2 &\propto& a + c_1 \ t_{p4}^{\rm HT} \ y_{p4} \;,
\label{apprx2}
 \\
 E_3 &\propto& b + c_2 \ t_{c}^{\rm HT}  \ y_{p4} 
                 + c_3 \ t_{p4}^{\rm HT} \ y_c \;, 
\label{apprx3}
\end{eqnarray}
where $a$, $b$ and $c_i$ ($i=1,2,3$) are independent of $y_\mu$, 
and $t^{\rm HT}_{p4}<0<t^{\rm HT}_{c}$. 
Note that we ignore $y_{\mu}^2$ in deriving the above equations, 
and the coefficients $c_i$ are positive 
under the condition $2V_c \geq V_p$. 
We can infer from Eq. (\ref{apprx2}) that $y_{p4}$ increases linearly 
as a function of $s_{p4}$ when the ground state is the HCO-$t_{p4}$ state. 
Because of the high symmetry with respect to $y_{p1} \leftrightarrow -y_{p1}$, 
Eqs. (\ref{apprx2}) and (\ref{apprx3}) are independent of $y_{p1}$. 
Once the distortions $y_{p4}$ and $y_c$ are substantially large, however, 
the distortion $y_{p1}$ is numerically found to stabilize 
the HCO-$t_{p4}$ further. 

Before closing this section, we make a brief comment on 
the spinless fermion case.\cite{CHotta}
In the $U\rightarrow\infty$ limit of the model(\ref{model}), 
the HCO state becomes unstable 
even if the modulations of the transfer integrals are introduced 
because the energy gain from spin fluctuations is absent. 
Thus we demonstrate that cooperative effects of electron 
correlations and electron-phonon couplings are important 
in the $\theta$-RbZn salt. 
The stability of the HCO state relative to the diagonal CO state suggested 
in the spinless fermion case without electron-phonon couplings\cite{CHotta} 
might be caused by the absence of phase factors, which are present if the 
$U\rightarrow\infty$ limit is naively taken. 
\section{Summary}
\label{summary}
We have investigated the cooperative effects of electron correlations 
and lattice distortions on the charge ordering in the $\theta$-RbZn salt.
By means of the exact-diagonalization method for systems with up to 16 sites, 
we have calculated the hole-hole correlation functions, the hole densities and 
the modulations of transfer integrals to clarify 
the role of electron-phonon couplings in this salt. 

In the absence of electron-phonon interactions, 
there appear three uniform phases characterized by 
quantum tunneling between the vertical-stripe COs, 
that between the diagonal-stripe COs, and 
that between the vertical- and the diagonal-stripe COs. 
It is found that all of these phases are changed into the horizontal-stripe CO 
by the introduction of electron-phonon couplings 
relevant to the $\theta$-RbZn salt. 
We can mostly reproduce the low-symmetry structure of $\theta$-RbZn 
at low temperature by using the model based on 
the high-symmetry structure at high temperature 
and by choosing the coupling strengths appropriately. 

We conclude that the structural deformation assists the horizontal-stripe CO 
that is experimentally found.\cite{Watanabe} 
In particular, the effects of molecular translations in the $c$-direction and 
molecular rotations are found to be stronger than 
that of translations in the $a$-direction. 
Thus, electron-phonon couplings are significant to stabilize 
the HCO-$t_{p4}$ and to realize the low-symmetry structure of $\theta$-RbZn. 
Otherwise the long-range Coulomb interactions favored a different CO pattern. 
With the help of the perturbation theory from the strong-coupling limit, 
we easily understand the mechanism for stabilizing the HCO-$t_{p4}$ state 
and the linearly increasing $y_{p4}$ and $y_c$ modulations. 
\begin{acknowledgments}
The authors are grateful to S. Iwai for showing his data prior to publication 
and Y. Yamashita for fruitful discussions.
This work was supported by the Next Generation SuperComputing Project, 
Nano Science Program, and 
Grants-in-Aid from the Ministry of Education, Culture, Sports, Science and 
Technology, Japan. 
\end{acknowledgments}
%
%


\begin{thebibliography}{99}
%
\bibitem{Ishiguro}
T. Ishiguro, K. Yamaji and G. Saito, 
{\it Organic Superconductors}
(Springer, Heidelberg, 1998). 
%
\bibitem{Williams}
J. M. Williams, J. R. Ferraro, R. J. Thorn, K. D. Carlson, U. Geiser, 
H. H. Wang, A. M. Kini and M.-H. Whangbo, 
{\it Organic Superconductors (Including Fullerenes: Synthesis, Structure, 
Properties, and Theory)}
(Prince Hall, Englewood Criffs, NJ, 1992). 
%
\bibitem{Shimizu}
Y. Shimizu, K. Miyagawa, K. Kanoda, M. Maesato and G. Saito, 
Phys. Rev. Lett. {\bf 91}, 107001 (2003). 
%
\bibitem{Kanoda}
K. Kanoda, 
Physica C {\bf 282}-{\bf 287}, 299 (1997). 
%
\bibitem{Kino}
M. Kino and H. Fukuyama, 
J. Phys. Soc. Jpn. {\bf 65}, 2158 (1996). 
%
\bibitem{Seo1}
H. Seo, 
J. Phys. Soc. Jpn. {\bf 69}, 805 (2000). 
%
\bibitem{Mazumdar}
S. Mazumdar, R. T. Clay and D. K. Campbell, 
Phys. Rev. B {\bf 62}, 13400 (2000). 
%
\bibitem{Clay}
R. T. Clay, S. Mazumdar and D. K. Campbell, 
J. Phys. Soc. Jpn {\bf 71}, 1816 (2002). 
%
\bibitem{SeoReview}
H. Seo, J. Merino, H. Yoshioka and M. Ogata, 
J. Phys. Soc. Jpn. {\bf 75}, 051009 (2006); and references therein. 
%
\bibitem{HMori1}
H. Mori, S. Tanaka and T. Mori, 
Phys. Rev. B {\bf 57}, 12023 (1998). 
%
\bibitem{HMori2}
H. Mori, S. Tanaka, T. Mori, A. Kobayashi and H. Kobayashi, 
Bull. Chem. Soc. Jpn {\bf 71}, 797 (1998). 
%
\bibitem{Watanabe}
M. Watanabe, Y. Noda, Y. Nogami and H. Mori, 
J. Phys. Soc. Jpn. {\bf 73}, 116 (2004). 
%
\bibitem{Miyagawa}
K. Miyagawa, A. Kawamoto and K. Kanoda, 
Phys. Rev. B {\bf 62}, R7679 (2000). 
%
\bibitem{Chiba}
R. Chiba, H. Yamamoto, K. Hiraki, T. Takahashi and T. Nakamura, 
J. Phys. Chem. Solids {\bf 62}, 389 (2001). 
%
\bibitem{HTajima}
H. Tajima, S. Kyoda, H. Mori and S. Tanaka, 
Synth. Met. {\bf 120}, 757 (2001). 
%
\bibitem{Yamamoto}
K. Yamamoto, K. Yakushi, K. Miyagawa, K. Kanoda and A. Kawamoto, 
Phys. Rev. B {\bf 65}, 85110 (2002). 
%
\bibitem{Iwai}
S. Iwai, K. Yamamoto, A. Kashiwazaki, F. Hiramatsu, H. Nakaya, Y. Kawakami, 
K. Yakushi, H. Okamoto, H. Mori and Y. Nishio, 
Phys, Rev. Lett. {\bf 98}, 097402 (2007)
%
\bibitem{Ducasse}
L. Ducasse, F. Fritsch and F. Caster, 
Synth. Met. {\bf 85}, 1627 (1997). 
%
\bibitem{Imamura}
Y. Imamura, S. Ten-no, K. Yonemitsu and Y. Tanimura, 
J. Chem. Phys. {\bf 111}, 5986 (1999). 
%
\bibitem{TMori1}
T. Mori, 
Bull. Chem. Soc. Jpn. {\bf 73}, 2243 (2000). 
%
\bibitem{TMori2}
T. Mori, 
Bull. Chem. Soc. Jpn. {\bf 71}, 2509 (1998); 
T. Mori, H. Mori, and S. Tanaka,
Bull. Chem. Soc. Jpn. {\bf 72}, 179 (1999); 
T. Mori,
Bull. Chem. Soc. Jpn. {\bf 72}, 2011 (1999). 
%
\bibitem{Merino}
J. Merino, H. Seo and M. Ogata, 
Phys. Rev. B {\bf 71}, 125111 (2005). 
%
\bibitem{TMori}
T. Mori, 
J. Phys. Soc. Jpn. {\bf 72}, 1469 (2003). 
%
\bibitem{EDagotto}
E. Dagotto, 
Rev. Mod. Phys. {\bf 66}, 763 (1994). 
%
\bibitem{HTajima2}
H. Tajima, S. Kyoden, H. Mori and S. Tanaka, 
Phys. Rev. B {\bf 62}, 9378 (2000). 
%
\bibitem{Tanaka}
Y. Tanaka and K. Yonemitsu, 
J. Phys. Soc. Jpn. {\bf 76}, No. 5 (2007). 
%
\bibitem{CHotta}
C. Hotta, N. Furukawa, A. Nakagawa and K. Kubo, 
J. Phys. Soc. Jpn. {\bf 75}, 123704 (2006). 
%
\end{thebibliography}

\end{document}